\documentclass[11pt]{article}
\usepackage[utf8]{inputenc}
\usepackage{amsmath,amsfonts,amssymb,latexsym}
\usepackage[T1]{fontenc}
\newtheorem{satz}{Theorem}[section]

\newtheorem{defi}[satz]{Definition}

\newtheorem{bem}[satz]{Remark}
\newtheorem{lemma}[satz]{Lemma}
\newtheorem{koro}[satz]{Corollary}
\newtheorem{conclusion}[satz]{Conclusion}
\newtheorem{ob}[satz]{Observation}

\newtheorem{conjecture}[satz]{Conjecture}

\newcommand{\mbb}{\mathbb}
\newcommand{\tit}{\textit}

\newcommand{\R}{\mathbb{R}}

\begin{document}
\thispagestyle{empty}
\begin{center}
\vspace*{1.0cm}
{\Large{\bf Gravitons as Goldstone Modes and the Spontaneous Symmetry Breaking of\\ Diffeomorphism Invariance}}
\vskip 1.5cm

{\large{\bf Manfred Requardt}}

\vskip 0.5cm

Institut fuer Theoretische Physik\\
Universitaet Goettingen\\
Friedrich-Hund-Platz 1\\
37077 Goettingen \quad Germany\\
(E-mail: requardt@theorie.physik.uni-goettingen.de)

\end{center}

\begin{abstract}
We complement investigations of the gauge structure of general relativity with an analysis of the physical consequences of the spontaneous breaking of diffeomorphism invariance which manifests itself in e.g. the socalled Einstein hole problem. We analyze the nature of the gravitons as Goldstone excitations both in the classical and the quantum case. We show that the metrical field and the classical space-time manifold play the role of an order parameter field and order parameter manifold as macroscopic super structures living in an underlying presumed quantum space-time. We furthermore relate our observations to possible phase transitions in some pre big-bang era.

\end{abstract} \newpage
\setcounter{page}{1}
\section{Introduction}
In classical general relativity (GR), in contrast to mathematical differential geometry, it is kind of a dogma that the points of the space-time manifold (S-T) have no real physical individuality. This was already realized in the context of the socalled \tit{Einstein-Hole-Argument} (see for example \cite{Pais},\cite{Stachel},\cite{Norton} or sect.4 in \cite{Rovelli}) and condensed in the mathematical statement
\begin{ob} In classical general relativity all diffeomorphic S-T-manifolds are physically indistinguishable. I.e., with $\Phi$ a diffeomorphism from $M$ to $M'$ (that is, not simply a coordinate transformation), and $g':=\Phi_{\ast}\circ g$ (note that $ \Phi_{\ast}$ denotes the push forward related to the pull back $((\Phi^{-1})^{\ast})$, $(M,g)$ and $(M',g')$ describe the same classical physics. In other words
\begin{equation}S\!-\!T=Riem/Diff        \end{equation}
\end{ob}
(cf.  e.g. \cite{Hawking})

In the special case $M'=M$ we would have a family of mathematically discernible metrics at the same (coordinate) point, that is
\begin{equation} g(x)\neq \Phi_{\ast}\circ g(x)  =g'(x)     \end{equation}
However one should note that, physically, we always have given a single metric from the family on the S-T-manifold as, by construction, the metrical properties on the S-T-manifold are given by a concrete measurement prescription which comes in a sense from outside in contrast to mathematics. We will come back to this topic below in the context of spontaneous symmetry breaking (SSB).

The typical way  how a metric is introduced in GR exploits the existence of local inertial frames (LIF) in which special relativity holds sway and which allow to perform the usual length- and time-measurements. The \tit{equivalence principle} and \tit{general covariance} then allow to transplant the respective measurement results into arbitrary coordinate systems.
\begin{conclusion}On a given S-T-manifold we can hypothetically envisage several mathematically different but physically equivalent metrical tensors
\begin{equation} g(x)\; ,\; g'(x)=\Phi_{\ast}\circ g(x)      \end{equation}
\end{conclusion}

We would like to emphasize that in our context of SSB we always have a large class of mathematically different but physically equivalent metrics (in the above sense) on $M$. Typically, SSB is concerned with the ground state or vacuum state of a system. In our context of GR and/or quantum gravity (QG) this means S-T devoid of macroscopic matter/energy content, i.e. solutions with vanishing \tit{Ricci-curvature}, that is
\begin{equation}R_{\mu\nu}=0\quad\text{or}\quad G_{\mu,\nu}=0    \end{equation}
These vacuum solutions have a large class of diffeomorphisms connecting them. Local deformations of the metric as in the \tit{hole-argument} will for example suffice.

While we will develop the subject matter from a direction starting with a view on SSB as it occurs in systems of many degrees of freedom (DoF) like e.g. condensed matter physics, there has been a chain of reasoning which exploits similarities between GR and classical gauge theories. 

A well written early paper, belonging to this class, is for example \cite{Isham}, in which relations between the \tit{tetrad formalism} of GR and socalled \tit{nonlinear realizations} of \tit{gauge groups} on \tit{coset spaces} in high energy physics are established (having been of some prominence at that time). We mention also \cite{Freund}. Written in a similar vein are papers from the Russian school (up to very recent times). To mention a few, \cite{Sarda1},\cite{Sarda2} or \cite{Sarda3}. This approach relies mainly on the fibre bundle framework of socalled \tit{gauge gravity theory} and the reduction theory of principal bundles with respect to the structure groups being used and is of a more formal character. A recent paper, also starting from this group-reduction point of view in connection with SSB is \cite{Tomb}. 

It is our aim in the following to unify this more formal group-theoretic approach with a different train of ideas which start from the more concretely given implications of SSB and \tit{gravitons} as \tit{Goldstone modes}, thus emphasizing features which may establish a connection to an underlying bundle of phenomena belonging to the not yet existing field of quantum gravity (QG). Note for example the remark in \cite{Isham}:
\begin{quote}
One does not expect any new development in the notoriously difficult problem of quantizing gravity to result from this modified point of view. However some insight may be gained\ldots
\end{quote}

It is our impression that the observation that gravitons are the Goldstone modes of SSB of diffeomorphism invariance will lead to real physical consequences if one can relate the more formal and abstract aspects on the level of classical gauge theory and fibre bundle reductions to the corresponding physical implications on the deeper levels of quantum space-time physics. 
\section{Physical Considerations concerning SSB of Diffeomorphism Invariance}
As the group theoretic aspects of SSB are represented in great detail in the above mentioned literature, we begin our analysis with the development of a more physical point of view concerning the subject matter which makes contact with related phenomena of SSB in systems consisting of many degrees of freedom.

There are two particular points to be mentioned which may shed some light on the scene in GR and QG. We try to elucidate them by briefly discussing two characteristic examples taken from the field of SSB and phase transitions in many-body physics. We will however only stress the points which are of relevance for our corresponding analysis in gravitational physics.

To begin with, we discuss the phenomenon of breaking of translation invariance by crystallization of a continous (quantum) many-body system. In the symmetric unbroken phase the particle density  $\rho (x):=<\hat{\rho}>$ is a constant, i.e.
\begin{equation}\rho (x+a)=\rho(x)\quad ,\quad x,a\in \mbb{R}^d    \end{equation}
Below some critical point or phase transition line we have instead a periodic dependence of the particle density in the respective \tit{pure phases}, i.e.
\begin{equation}\rho (x+a)\neq\rho(x)\quad\text{in general}      \end{equation}
 but
\begin{equation}\rho (x+R_i)=\rho(x)   \end{equation}
for some discrete subgroup of $\mbb{R}^d$.

\begin{defi}By a pure phase we mean in the above context a crystal having a definite macroscopic position in space. Mixtures may occur if we average over a group of such localized crystals. In a pure phase correlation functions do decay but only slowly due to the existence of collective Goldstone excitations.
\end{defi}
\begin{bem}A method of generating such localized crystal is the method of Bogoliubov quasi-averages (an external localizing field which is switched off in the end after the thermodynamic limit has been taken).
\end{bem} 

\begin{ob}If this happens, both in the classical and the quantum regime, long-lived collective excitations do emerge which induce long-range correlations. In the case of a crystal they are called phonons.
\end{ob}
Representations of the Goldstone phenomenon in the quantum regime are so numerous that we mention only very few sources. Almost every textbook about quantum field theory  contains a brief discussion (see e.g. \cite{Itzykson}). As to the older literature there is the nice comprehensive review \cite{Guralnik}. A more recent contribution is for example \cite{Requ2}. A detailed development of the Goldstone phenomenon in the regime of classical statistical mechanics can be found in \cite{Requ1}.

There exists an important difference between Goldstone particles in say QFT and in e.g. condensed matter physics and statistical mechanics. In QFT the Goldstone particles are exact mass -zero particles, i.e. they have a sharp excitation branch. On the other hand, in condensed matter physics, or systems having a non-vanishing particle density in general, they aquire an infinite lifetime only for momentum zero, while for non-vanishing momenta they are usually still relatively stable collective excitations but have only a finite lifetime (which typically decreases with increasing momentum) resulting in a smeared dispersion law (cf. e.g. \cite{Requ2}). Furthermore, while in RQFT their spin vanishes due to general principles (see \cite{Reeh}), this is not so in the more general context. The underlying reason is the absence of Lorentz covariance, Einstein causality and the socalled spectrum condition ( energy-momentum concentrated in the forward cone). Furthermore, while in most scenarios we can at least exploit translation invariance and the corresponding Fourier-mode decomposition, this is absent in GR and QG. Therefore, in the following, we will avoid all these concepts and discuss the Goldstone phenomenon in a much broader framework. 

The relevant point in our investigation will be the following:
\begin{ob}For a hypothetical observer living inside one of the respective pure phases, i.e. the crystal, being translated by some vector, $a$, the internal physics is the same compared to a corresponding observer in a crystal,  being translated by some vector, $a'\neq a$, provided corresponding coordinate systems have been chosen. Only an outside observer is able to discern the various translated pure phases.
\end{ob}
For illustrational purposes we mention another example, i.e. a lattice spin system, being capable of spontaneous magnetization. A pure phase in this scenario is described by a magnetization vector, pointing in a certain direction in configuration space. Again, the internal physics relative to the orientation of this  magnetization vector is the same in all the different pure phases. Only an external observer is able to see the different phases (that is, the different directions of magnetization) by using his external reference system.
\begin{bem}All the internal observers are however able to observe the long-range collective Goldstone excitations, that is, phonons or magnons.
\end{bem}

All this now winds up to the observation that all internal observers see essentially the same physics provided they adapt their internal reference systems appropriately. That is, the situation is completely the same as compared to the case of diffeomorphism invariance of vacuum solutions in GR or QG. 
\begin{conclusion}By the above observations we feel entitled to attribute to the different members of the class of diffeomorphic realizations of S-T a perhaps less than ephemeral or only formal existence as, by necessity, we are only internal observers in the latter case.
\end{conclusion}

We want to conclude this section with a brief analysis of the character of the goldstone modes under discussion. Phonons are essentially lattice vibrations in the crystal case, magnons are fluctuations of the local magnetization. Phrasing it somewhat differently one can venture to say:
\begin{ob}The Goldstone modes try to locally interpolate between the different potentially coexisting pure phases. I.e., local distortions of the crystal lattice can for example be regarded as local transitions into another slightly shifted crystal configuration. The same holds in the magnon case.
\end{ob}
\begin{conclusion}Exploiting the above correspondence between our examples and diffeomorphism invariance of S-T, one may conclude that the Goldstone modes in the latter case are the gravitons, acting as local distortions of S-T. They interpolate locally between the mathematically different but physically only hypothetically coexisting realizations of S-T as we are living in only one of these possible realizations.
\end{conclusion}    
\section{The Conceptual Representation of SSB of Diffeomorphism Invariance in the Context of General Relativity and Quantum Gravity}
In this section we want to analyse the nature of SSB in our context. Note that the different diffeomorphic realisations of S-T can be viewed as an underlying differentiable manifold being equipped with different but diffeomorphic pseudoriemannian metrics. Furthermore, as we are mainly interested in the case of degenerate ground states (in a possibly underlying theory of QG), we assume that the (macroscopic) energy-momentum tensor vanishes.

The cornerstone of GR is the \tit{equivalence principle}, that is, at every point, $P$, of the S-T manifold there exists for a fixed metrical field, $g(\circ,\circ)$, a class of LIF in which the laws of special relativity (SR) hold in an at least infinitesimal neighborhood of the point $P$. Mathematically we can construct such a local coordinate system as follows, while we take at the same time the opportunity to introduce a number of useful concepts and notations (cf. e.g. \cite{Moeller} sect. 9.6. As to the tetrad formalism see also \cite{Synge} sect. I,3 ). 
\begin{defi}In a given coordinate system, $x$, a (contravariant) tetrad at $P$ is given by 4 pseudoorthogonal tangent vectors, $e_a=(e_a^{\nu})$ with $a=0,1,2,3$ labelling the 4 vectors and $\nu$ denoting the indices with respect to the local coordinate tangent vectors $\partial_0,\partial_1,\partial_2,\partial_3$. We have 
\begin{equation}e_{a\nu}=g_{\nu\mu}e_a^{\mu}\quad ,\quad e_a^{\nu}e_{b\nu}=\eta_{ab} =g(e_a,e_b)    \end{equation}
with $\eta_{ab}$ the Minkowski tensor.
\end{defi}

For formal reasons we introduce
\begin{equation}e^{a\nu}:= \eta^ {ab}e_b^{\nu}\quad , \quad e_{\nu}^a:= \eta^ {ab}e_{b\nu}               \end{equation}
\begin{ob}
\begin{equation}e_a^{\nu}e^b_{\nu}=\delta_a^b            \end{equation}
\end{ob}
\begin{lemma}
\begin{equation}e_a^{\nu}e_{\mu}^a=\delta_{\mu}^{\nu}          \end{equation}
\end{lemma}
This follows from the preceding observation. I.e., we have
\begin{equation}(e_a^{\nu}e_{\mu}^a)e_b^{\mu}= e_a^{\nu}(e_{\mu}^ae_b^{\mu})=e_a^{\nu}\delta_b^a=e_b^{\nu}=\delta_{\mu}^{\nu}e_b^{\mu}        \end{equation}
\begin{bem}Note that $\nu,\mu$ refer to the covariant coordinate indices and are consequently raised and lowered with the help of $g_{\nu\mu},g^{\nu\mu}$ while $a,b$ as formal indices are raised and lowered with the help of the Minkowski metric.
\end{bem}

\begin{ob}Any two tetrades, $(e_a),(f_b)$, at $P$ are connected by a Lorentz transformation, $L$, i.e.
\begin{equation}e_a^{\nu}=L_a^{\cdot b}f_b^{\nu}\quad\text{or}\quad e_a=L_a^{\cdot b}f_b       \end{equation}
\end{ob}
\begin{lemma}
\begin{equation}e_{\nu}^af_b^{\nu}=L_{\cdot b}^a       \end{equation}
with
\begin{equation}e_{\nu}^a=L_{\cdot b}^a f_{\nu}^b\quad ,\quad f_b^{\nu}=L_{\cdot b}^a e_a^{\nu}     \end{equation}
\end{lemma}
This follows from
\begin{equation}(e_{\nu}^af_b^{\nu})f_{\mu}^b=e_{\nu}^a(f_b^{\nu}f_{\mu}^b)=e_{\nu}^a\delta_{\mu}^{\nu}=e_{\mu}^a    \end{equation}
and
\begin{equation}e^a=L_{\cdot b}^af^b\end{equation}
\begin{lemma}We have
\begin{equation}\eta_{ab}=g(e_a,e_b)=g(f_a,f_b)              \end{equation}
\end{lemma}
Proof: Under the assumption $g(f^c,f^d)=\eta^{cd}$ we have
\begin{equation}g(e^a,e^b)=g^{\nu\mu}e_{\nu}^ae_{\mu}^b= g^{\nu\mu}f_{\nu}^cf_{\mu}^dL^a_{\cdot c}L^b_{\cdot d}=\eta^{cd}          L^a_{\cdot c}L^b_{\cdot d}=\eta^{ab}          \end{equation}

As shown in \cite{Moeller}, l.c., one can easily construct a new local coordinate system with the help of the tetrad at $P$:
\begin{equation}(x')^i:=e_{\nu}^i(x^{\nu}-x_P^{\nu})\quad ,\quad x^{\nu}=x_P^{\nu}+e_k^{\nu}(x')^k   \end{equation}
which is pseudo-orthogonal at $P$, i.e.
\begin{equation}g'_{ik}(P)=e_i^{\nu}e_k^{\mu}g_{\nu\mu}=\eta_{ik}  \end{equation}
Other pseudo-orthogonal coordinate systems can be generated by Lorentz transformations from the $(e^a)$-system, i.e.
\begin{equation}f_a^{\nu}=L_a^{\cdot b}e_b^{\nu}\quad\text{and}\quad y^i=f_{\nu}^i(x^{\nu}-x_P^{\nu})   \end{equation}
It follows
\begin{equation}y^i=L_{\cdot k}^i(x')^k   \end{equation}

However, a pseudo-orthogonal coordinate system at $P$ is in general not a LIF. Therefore, an observer at $P$ in such a system still experiences a gravitational field. This can be transformed away by means of a more general coordinate transformation which leads to the
\begin{ob}There exists a coordinate transformation to socalled local Lorentz-coordinates (e.g. Riemann normal coordinates) such that 
\begin{equation}g_{\nu\mu}(P)=\eta_{\nu\mu}\quad\text{and}\quad \partial  g_{\nu\mu}(P)/\partial x_{\rho}=0\quad\text{or}\quad  \Gamma_{\nu\mu}^{\rho}(P)=0      \end{equation}
which is the definition of a LIF.
Further local Lorentz transformations leave the class of LIF invariant while leading, of course, to another system of local Lorentz-coordinates.
\end{ob}

We now will establish the connection to SSB of diffeomorphism invariance. Central in the context of SSB and phase transitions is the notion of \tit{order parameter}. In the above examples, taken from many-body theory, order parameters are certain observable quantities which characterize the pure spontaneously broken phases as e.g. the gradient of the particle density or the magnetization. One can equally well use the corresponding quantum observables, that is, observables whose ground state expectation values vanish in the ordered unbroken phase while being different from zero in the broken phases. Furthermore, the different ground state expectation values are connected by the set of broken symmetry transformations.

More specifically, we have a large group, $G$, of symmetrie (some of which are broken) containing a closed subgroup, $H$, of conserved symmetries. The \tit{configuration manifold} of ground states can be related to and parametrized by the \tit{cosets} of the \tit{homogeneous space} $G/H$. We conclude, that it is important to study the structure of the space $G/H$ with $G$ a Lie group and $H$ a closed subgroup. This problem is in general not trivial and we will deal with the more mathematical aspects of the problem  in the following section.

We will now transplant this picture into the more general framework of GR and/or QG. As configuration manifold (of minima of some functional of the occurring fields) we take the set of vacuum solutions of GR. The group $G'$ is the (in general) infinite dimensional group $Diff$. In the following we restrict ourselves, for convenience, to a single orbit under $Diff$, that is, a fixed vacuum solution together with its transforms under $Diff$. In a next step we will shift our point of view to a more local one, i.e. we switch to the action of the respective groups in the corresponding \tit{tangent bundle} of S-T.
\begin{ob}Locally the group $Diff$ is realized by the group $G:=GL(n.\R)$ of general linear transformations in the respective local coordinate systems:
\begin{equation}Diff\:\rightarrow\: (\partial (x')^i/\partial x^j)(P)\: \in GL(n,\R)                 \end{equation}
\end{ob}

One can equally well consider the local action on the (principal) \tit{frame bundle} over a fixed manifold $M$, i.e. the action of $Diff$ in a local trivialization of the frame bundle.
\begin{defi}As a closed subgroup of $GL(n.\R)$ we take the orthochronous Lorentz group, $L$, (the connected component of the group unit, $e$).
\end{defi}
\begin{bem}In a given coordinate system the local frames are in a one-one relation to elements of $GL(n,\R)$, i.e.
\begin{equation}v_i=v_i^k(\partial/\partial x^k)\quad , \quad i=1,\ldots ,n        \end{equation}
with $(v_i^k)\in GL(n.\R)$.
\end{bem}

The field on $M$ we are mainly interested in is the classical metric field $g_{\nu\mu}(x)$ or 
\begin{equation}g_{\nu\mu}(x):=< \hat{g}_{\nu\mu}(x)>       \end{equation}
with $\hat{g}$ its quantum version (in some framework of semiclassical QG).
\begin{defi}We call  $g_{\nu\mu}(x)$ an order parameter field, thus generalizing the notion of order parameter. An order parameter manifold, $(M,g)$, is $M$ plus a fixed metrical field  $g_{\nu\mu}(x)$.
\end{defi}
\begin{ob}In a given fixed coordinate patch the group $Diff$ acts on the metrical field $g_{\nu\mu}(x)$ by acting at each point $P$ via $GL(n,\R)$ (in a tensorial way)
\begin{equation}Diff:\: g(x)\:\rightarrow\:g'(x)       \end{equation}
\end{ob}

The \tit{hole-problem} was solved by Einstein (cf. \cite{Rovelli} l.c.) by introducing for example four particles $A,B,C,D$ with $A,B$ meeting in point $x_i$ inside the hole and $C,D$ in point $x_j$. He observed that the diffeomorphism now transforms both the metric tensor $g(x)$ and the trajectories of the test particles so that they get shifted as well with their distance becoming equal with respect to the transformed metric compared to their distance relative to the original metric $g(x)$.
\begin{ob} This is exactly the situation we discussed in section 2 in the context of e.g. many-body physics. I.e., physically the different phases become indistinguishable if the respective reference frames are appropriately reoriented. In the hole argument the reference system is provided by the 4 particle trajectories and their intersection points. Put differently, the Einstein hole argument is in fact an illustration of our concept of SSB in general relativity.
\end{ob}

As a last point in this section we discuss what a local observer in a LIF physically experiences concerning the SSB of diffeomorphism invariance. We again emphasize that we have one metric field, $g(x)$, with the other potentially existing fields $g':=\Phi_*\circ g$ being unobservable for the local observer. What is actually at his disposal are \tit{passive} coordinate transformations. He can for example apply Lorentz transformations.
\begin{ob}We learned above that with $(e_a)$ a Lorentz-orthonormal frame, $g(e_a,e_b)=\eta_{ab}$, the same holds for the Lorentz transformed tetrad, $(f_a)$, i.e. $g(f_a,f_b)=\eta_{ab}$.
\end{ob}
That is, if the local observer uses a LIF (local Lorentz coordinates) he observes that gravitational forces are locally absent. The same holds then when he applies the Lorentz group $L_+^{\uparrow}$ (and extending the new frame appropriately to a local Lorentz coordinate system). If, on the other hand, he applies general elements from $GL(n,\R)$ he will observe gravitational forces in the transformed coordinate frame.
\begin{conclusion}In our view this is the local manifestation of the breaking of diffeomorphism invariance.
\end{conclusion}
In the next section we will relate our physical observations with the more abstract formalism of \tit{reduction of pricipal bundles}.
\section{SSB and Reducible Principal Bundles}
A large part of the above mentioned literature is concerned with the reducibility of the \tit{frame bundle} with \tit{structure group} $GL(n,\R)$ to a subgroup, in our case the Lorentz group $L_+^{\uparrow}$ or $SO(n-1,1;\R)$. Frequently this phenomenon is already considered as a case of SSB. In our view reducibility as such is a widespread phenomenon in the field of principal bundles. So we regard it rather as a necessary prerequisite for SSB, not as the phenomenon itself! The mathematical results we are using in the follwing can be found in the following books, \cite{KN},\cite{CB},\cite{CH},\cite{H}. We take some pains of mentioning the places where the results (and more) can be found because exact citations of the sometimes quite nontrivial results are frequently missing in the physical literature. 

We denote a general bundle by $(E,B,G,F)$ or simply $E$, $E\rightarrow B$. $E$ is the total space, $B$ the base space (typically in our case the space-time manifold $S-T$), $G$ is the structure group (in physics called the gauge group), $F$ the typical fibre (sometimes $F$ is a vector space). Both $E,B$ are differentiable manifolds. Points in $E,B$ are denoted by $p,x$. $E$ is continuously and surjectively projected by $\pi$ on $B$. With $x\in B$, $\pi^{-1}(x)=E(x)$ or $E_x$ is the fibre over $x$. $G$ acts on $F$ from the left via  homeomorphisms or (in case of a vector space) as linear automorphisms. All fibre bundles are assumed to be locally trivial, that is, it exists a covering of $B$ by open sets, $U_i$, so that it exists $\phi_i$ with 
\begin{equation}\phi_i:\pi^{-1}(U_i)\simeq U_i\times F          \end{equation}
In particular
\begin{equation}\phi_i':\pi^{-1}(x)=E_x\simeq F          \end{equation}
with
\begin{equation}\phi_i(p)=(\pi(p),\phi_i'(p)        \end{equation}
If $x\in U_i\cap U_j$ the transition maps
\begin{equation}\phi_i'\circ\phi_j'^{-1}:F\rightarrow F      \end{equation}
are elements of the structure group.
\begin{defi}A principal bundle, $P$, is a fibre bundle in which the typical fibre $F$ and the structure group $G$ are identical or, equivalently, $E_x$ or $P_x$ are diffeomorphic to $G$ (as Lie groups).
\end{defi}
\begin{ob}It is important that one can define a right action of $G$ on the fibres, $P_x$, i.e.
\begin{equation}R_g\circ p=p\cdot g\quad\text{or}\quad R_g\circ p=\phi_i^{-1}\circ(\phi_i(p)\circ g)    \end{equation}
under which $G$ acts freely on $P_x$.
\end{ob}
\begin{bem}Note that left action is in general not! fibre preserving (i.e. independent of the choice of patch $U_i$).
\end{bem}

In our context we are typically concerned with the bundle of frames, $L(M)$, over a manifold $M$. A linear frame at $x\in M$ is an ordered basis of the tangent space $T_x(M)$. The linear group acts on $L(M)$ from the right in the following way.
\begin{equation}Y_j=A_j^ix_i   \end{equation}
$(Y_j),(X_i)$ linear frames at $ x\in M$. Taking a coordinate basis $(\partial_{x^i})$ each frame at $x$ can be expressed as
\begin{equation}X_j=X^i_j \partial_{x^i}     \end{equation}
for some $(X^i_j)\in GL(n,\R)$. This shows at the same time the local triviality of the frame bundle, i.e.:
\begin{equation}\pi^{-1}(U)\simeq U\times GL(n,\R)    \end{equation}
\begin{defi}Let $P(M,G)$ be a principal bundle. Let $P'$ be a submanifold of $P$ and $H$ a Lie-subgroup of $G$ so that $P'(M,H)$ is again a principal bundle with structure group $H$. We call $P'(M,H)$ a reduction of $P(M,G)$. We say, the structure group $G$ is reduced to $H\subset G$.
\end{defi}
\begin{bem}This property is a nontrivial! result. It has a variety of interesting applications (cf. e.g. \cite{KN} or \cite{CB}).
\end{bem} 

In the reduction process an important role is played by the \tit{coset space} $G/H$, or in our contex, $GL(n,\R)/L_+^{\uparrow}$. In most of the mathematical literature it is dealt with the case $GL(n,\R)/O(n,\R)$, i.e. $O(n,\R)$ instead of $L_+^{\uparrow}$. The latter situation is technically much simpler (as is the case for Riemannian geometry in general compared to Lorentzian geometry) but one finds frequently the slightly erroneous statement in the physical literature that the former case is more or less equivalent. Therefore we want to briefly comment on this point.

We have the general result (cf. \cite{KN} p.43 or \cite{CB} p.109)
\begin{satz}$G/H$ is an analytic manifold , in particular the projection $G\rightarrow G/H$ is real analytic. Furthermore, $G$ is locally diffeomorphic to $G/H\times H$, i.e. $G(G/H,H)$ with $H$ as structure group is a principal bundle (cf. \cite{KN} p.55).
\end{satz} 
In special cases, e.g. if $H$ is a maximal compact subgroup, stronger results hold. 
\begin{satz}With $H$ a maximal compact subgroup
\begin{equation}G\simeq H\times \R^m         \end{equation}
\end{satz}
This is a result by Iwasara (cf. \cite{CB} p.386 or\cite{CH} p.109ff.
\begin{koro}We have 
\begin{equation}GL(n,\R)/O(n)\simeq \R^{n(n+1)/2}      \end{equation}
with $dim(O(n))=n(n-1)/2$.
\end{koro}
\begin{bem}One should note that $L_+^{\uparrow}$ is not compact and (to our knowledge) a corresponding result does not hold for $L_+^{\uparrow}$ instead of $O(n)$.
\end{bem}

The reason why such a result holds for $O(n)$ can be understood relatively easily. The wellknown \tit{polar decomposition} tells us that there exists an essentially unique (global!) decomposition. 
\begin{satz}[Polar Decomposition]It exists an essentially unique decomposition
\begin{equation}L=O\cdot |L|      \end{equation}
with $O$ orthogonal and $L\in GL(n,\R)$, $|L|$ positive semidefinite, more specifically 
\begin{equation}|L|=(L^+\cdot L)^{1/2}     \end{equation}
and $L^+$ the adjoint of $L$.
\end{satz}
\begin{bem}This important result is much more general and holds also for \tit{closable (unbounded) operators} (cf. e.g. \cite{RS} vol.I)
\end{bem}
It is crucial for such a result to hold that one can exploit the spectral theorem, i.e. that for example $|L|$ is well-defined and, a fortiori, selfadjoint. Nothing in that direction holds (to our knowledge) if  $O$ is replaced by an element of $L_+^{\uparrow}$. In the latter case one only has a weaker result of a (in our view) quite different nature (which is however sufficient in our case).

With $G$ a Lie group, $H$ a closed subgroup, let $\hat{G},\hat{H}$ be the respective Lie algebras, $\hat{M}$ some vector subspace of $\hat{G}$ so that   $\hat{G}$ is the direct sum, $\hat{G}=\hat{M}+\hat{H}$, (note that $\hat{M}$ is not! unique in general). Let $\exp_{\hat{M}}$ be the restriction of the exponential map to $\hat{M}$ and $\hat{e}:=e\cdot H=H$ the element in $G/H$ under the projection $\pi:G\rightarrow G/H$.
\begin{satz}(cf. e.g. \cite{H} p.113) There exists a neighborhood, $U$, of $0$ in $\hat{M}$ which is mapped homeomorphically under $\exp_{\hat{M}}$ onto $\exp_{\hat{M}}(U)\subset G$ so that $\pi$ maps $\exp_{\hat{M}}(U)$ homeomorphically onto a neighborhood of $\hat{e}$ in $G/H$.
\end{satz}
To prove this important theorem, one needs the following lemma (\cite{H} p.105), which is useful in various contexts.
\begin{lemma}With $\hat{G}=\hat{M}+\hat{H}$ there exist open neighborhoods $U_{\hat{M}},U_{\hat{H}}$ of $0$ in $\hat{M},\hat{H}$ so that the map
\begin{equation}(A,B)\rightarrow \exp(A)\cdot\exp(B)      \end{equation}
is a diffeomorphism of $U_{\hat{M}}\times U_{\hat{H}}$ onto an open neighborhood of $e$ in $G$.
\end{lemma}
\begin{bem}It is remarkable that this map generates a full neighborhood of $e$ in $G$ as it is, at first glance, only a product set. It follows from the particular situation in Lie groups. It hinges in particular on the non-vanishing of a certain functional determinant around $e$ in $G$. This guarantees the bijectivity of the map near $e$. But one does not know how large this neighborhood actually is.
\end{bem}

\begin{conclusion}The above theorem says essentially that with the help of $\hat{M}$ we get locally! a transversial submanifold relative to $H$ which yields a parametrization of the fibres around $\hat{e}\in G/H$. Via group multiplication we can then transport this parametrization to a neighborhood of any coset $g\cdot H$.
\end{conclusion}
\begin{satz}In the case of $GL(n,\R),L_+^{\uparrow}(n,\R)$, matrices with 
\begin{equation}\eta\cdot A^T=A\cdot\eta \quad\text{(pseudosymmetric)}        \end{equation}  
span a $n(n+1)/2$-dimensional submanifold being locally transversal to $L_+^{\uparrow}(n,\R)$, i.e. they locally coordinatize $GL(n,\R)/L_+^{\uparrow}(n,\R)$. That is, in a neighborhood of $e\in GL(n,\R)$ each element $L\in  GL(n,\R)$ can be written uniquely as
\begin{equation}L=\Lambda\cdot A       \end{equation}  
with $\Lambda\in L_+^{\uparrow}(n,\R)$ and $A$ pseudosymmetric.
\end{satz}
\begin{bem}Note that e.g. matrices with $A^T=A$ do not have this property while also spanning a $n(n+1)/2$-dimensional submanifold. There exist Lorentz boosts being even positive!. The above local decomposition with $A$ pseudosymmetric was used in \cite{Isham}.
\end{bem}
Proof of theorem: We show that no element in $L_+^{\uparrow}$ different from $e$ is pseudosymmetric. We have with $\Lambda\in L_+^{\uparrow}$:
\begin{equation}\Lambda^T\eta\Lambda=\eta     \end{equation}
Assume that
\begin{equation}\Lambda^T=\eta\Lambda\eta    \end{equation}
we get
\begin{equation}\eta\Lambda\eta^2\Lambda=\eta\quad\rightarrow\quad \eta\Lambda^2=\eta        \end{equation}
hence 
\begin{equation}\Lambda^2=e\quad\rightarrow\quad\Lambda=e       \end{equation}
which proves the theorem as it holds $n(n-1)/2+n(n+1)/2=n^2= Dim(GL(n,\R)$.

There is an important theorem in the reduction theory of principal bundles (\cite{KN} p.57) saying the following:
\begin{satz}The principal bundle $P(M,G)$ is reducible to $P'(M,H)$ iff $P/H$ admits a cross section.
\end{satz}
\begin{bem}The meaning of $P/H$ is the following. The fibres of $P$ are diffeomorphic to $G$, hence the fibres of the bundle $P/H$ are diffeomorphic to the typical fibre $G/H$. The structure group is again $G$ (multiplication from the left).
\end{bem}
In the case of $O(n)$ 
\begin{equation}GL(n,\R)/O(n)\simeq\R^{n(n+1)/2}    \end{equation}
it is easy to see that local cross sections can be pasted together and extended to a global cross section (cf. \cite{CB} p.385) because $\R^m$ is a vector space and assuming that $M$ is paracompact.
\begin{ob}In the case of $L_+^{\uparrow}$ as subgroup we have local cross sections around every point of $P/H$ or rather $L(M)/L_+^{\uparrow}$ ($L(M)$ the frame bundle) as a consequence of our above results. But (to our knowledge) $GL(n,\R)/L_+^{\uparrow}$ is not diffeomorphic to a vector space.
\end{ob}

We can however proceed as follows. We assume that a Lorentzian metric, $g$, is given on our space-time manifold $M=S-T$.
\begin{bem}In contrast to the Riemannian case (paracompact manifold) this is not automatic (cf. e.g. \cite{CB} p.293).
\end{bem}
With the help of $g$ we generate the set of pseudoorthogonal tetrads $(e_a)$ at every point $x$ of $M$. This set is invariant under $L_+^{\uparrow}$ as we have seen in section 3.
\begin{ob}The subset of $L(M)$ consisting of pseudoorthogonal tetrads at each $x\in M$ is a \tit{reduced subbundle} $Q(M)\subset L(M)$ with structure group $L_+^{\uparrow}$.
\end{ob}
In each fibre of $Q(M)$ $L_+^{\uparrow}$ acts freely from the right. Therefore the projection
\begin{equation}pr:L(M)\rightarrow L(M)/L_+^{\uparrow}    \end{equation}
is constant on the fibres of $Q$ as subsets of the fibres of $L(M)$.
\begin{ob}Thus $pr$ induces a mapping 
\begin{equation}s:M\rightarrow L(M)/L_+^{\uparrow}    \end{equation}
via
\begin{equation}s(x):=pr(I(e_a(x))    \end{equation}
with $\pi_Q(e_a(x))=x$ the projection map in the subbundle $Q(M)$ and $I$ the injection (imbedding) map 
\begin{equation}I:Q(M)\rightarrow L(M)    \end{equation}
It is obvious that $s(x)$ determines a cross section in $L(M)/L_+^{\uparrow}$.
\end{ob}
\begin{bem}Note that $s(x)$ represents the equivalence class of Lorentz frames in $L(M)/L_+^{\uparrow}$ over $x$. That this is a well defined mapping is obvious. The continuity or differentiability of the cross section is the crucial point. This follows from the properties of the composition of the above maps all of which are differentiable. Furthermore, with the help of this cross section we get back exactly the subbundle $Q(X)$  we started from.
\end{bem}
\section{Conclusion}
We have learned in the preceeding sections that the gravitational field (or, rather, the metrical field), $g$, can be regarded as an \tit{orderparameter field} and the macroscopic, smooth space-time manifold $(S-T;g)$ as an \tit{order parameter manifold}, lying above a presumed microscopic (irregular and erratic) quantum space-time consisting of an array of many interacting DoF.

We have invoked the situation of quantum many-body systems for several times. There the situation is the following. We usually have a symmetric unbroken phase with the order parameter being zero, and, in another region of certain parameters, a phase transition to a set of physically more or less identical spontaneously broken phases characterized by certain non-vanishing values of the order parameter. 

Transferring these observations to our field we may venture to say:
\begin{conjecture}We associate the presumed unbroken phase of our (quantum) space-time, that is, the absence of a classical, macroscopic metrical field, $g$, to some \tit{pre-big-bang era}. The emergence of classical space-time is then the result of a phase transition (which may have happened before or near the big-bang).
\end{conjecture}
 
Another interesting point concerns the nature of the gravitons themselves. In many-body physics the corresponding excitations are the phonons with the ordered phase being the crystal phase and the relatively long-lived lattice phonons as Goldstone particles. On the other hand, phonons do already exist in quantum fluids but they happen to be more strongly damped. The phonons which occur e.g. in fluids can be associated with the SSB of Galilei invariance (cf. e.g. \cite{Requ2}.
\begin{conjecture}We think that in our context gravitons, while being more stable in the ordered phase, i.e. $S-T$ plus a non-vanishing $g$-field, have already existed in the unordered phase (quantum vacuum with vanishing classical $g$). In this phase they represented certain types of more primordial excitations not being related to deformatons of classical $S-T$. They perhaps do reflect, as in the above Galilei-case, the existence of a quantum vacuum as such.
\end{conjecture}
\begin{bem}This may be interesting in connection with string theory. While string theory, at least as a starting point, exploits a classical embedding or target space, gravitons are associated with certain low-lying excitations of closed strings. Our above observations concerning the possible nature of gravitons may perhaps shed some light on a certain connection to string theory and the role of gravitons in this framework.
\end{bem} 

The last point we want to address is whether there exists an objective correlate to the different hypothetical phases being described by the different diffeomorphic metrical fields. In SSB of many-body systems all the different phases can really exist while only one is realized in each case. In the corresponding situation of GR we are less accustomed to such a picture. However, in more recent times the general perspective has changed a little bit, given the discussion about the \tit{landscape} in string theory and cosmology or \tit{induced/entropic gravity} which employ more or less openly some microscopic baclground substrate as supposed carrier of the concepts of physics.

\end{document}